# Incommensurate modulations in stoichiometric Ni$_2$MnGa ferromagnetic shape memory alloy: An overview

Sanjay Singh[1,2], S. R. Barman[1] and D. Pandey[3*]

[1]UGC-DAE Consortium for Scientific Research, Khandwa Road, Indore, 452001, India.

[2]Experimentalphysik, University Duisburg- Essen, D-47048 Duisburg, Germany.

[3]School of Materials Science and Technology, Institute of Technology, Banaras Hindu University, Varanasi-221005, India.

## Abstract

This article presents a brief overview of our recent work on the nature of long period modulation in the premartensite and martensite phases of Ni$_2$MnGa ferromagnetic shape memory alloy using high resolution synchrotron x-ray powder diffraction patterns. The commensurate structure model using the Pnnm space group is unable to account for the peak positions of the satellite reflections that appear due to modulations correctly. LeBail and Rietveld refinements using the (3+1)-D super space group *Immm(00γ)s00* show that the peak positions of all the reflections, including the satellites, can be explained satisfactorily using incommensurate modulations for both the premartensite and martensite phases. The incommensurate modulation vectors are found to be $\boldsymbol{q}$= 0.33761(5)$\boldsymbol{c}*$ = (1/3+δ$_1$)$\boldsymbol{c}*$ and 0.43160(3)$\boldsymbol{c}*$= (3/7+δ$_2$)$\boldsymbol{c}*$, where δ$_1$ and δ$_2$ are the degrees of incommensuration for the premartensite and martensite phases, respectively. The periodicity of the closest rational approximant of the premartensite and martensite phases are confirmed to be 3M and 7M, respectively, in agreement with single crystal diffraction results.

## 1. Introduction:

Phase transitions in Ni-Mn-Ga ferromagnetic shape memory alloys (FSMA) have received enormous interest because they exhibit very large magnetic field induced strain (MFIS). [1, 2, 3] These alloys are technologically more promising for actuator devices as compared to the conventional shape memory alloys (SMAs) as well as other materials like magnetostrictive Terfenol-D or piezoelectric ceramics [4], which exhibit strains of about 0.1% only as against 10% in FSMAs. The basis of shape change in both SMAs and FSMAs is a martensitic phase transition. However, the external stimulus for shape change in the martensite phase of FSMA's is an external magnetic field where as it is external stress or temperature in the conventional SMAs. The martensite transition is a diffusionless displacive type structural phase transition in which the atoms move in a cooperative way with respect to their neighboring atoms. [5] The martensitic transitions involve not only a lattice variant deformation, usually called as Bain distortion, of the high temperature austenite phase leading to the low temperature martensite phase but also a lattice invariant deformation that leaves the austenite-martensite interface (habit plane) undistorted and unrotated in an average sense at the microscopic scale by the formation of symmetry permitted martensite variants or ferroelastic domains through twinning or faulting. [6] When the conventional SMAs are deformed by external stress in the martensite state, the martensite variants merge and lead to a shape change and produce net strain. The deformed sample regains its original shape on heating above the martensitic transition temperature ($T_M$) in the austenite phase. In FSMAs, the shape change occurs within the martensite phase field by application of an external magnetic field that leads to merger of the different orientation variants of the martensite. This process results in large shape change but when the field is removed, the sample does not regain its original shape within the martensite phase unless an external stress is applied in a direction perpendicular to the magnetic field direction [7,8]. However, there is an upper limit of externally applied stress above which magnetic-field-induced strains are completely suppressed and this limiting value of stress is known as blocking stress[7,8]. FSMAs are more suitable for practical applications than conventional shape memory alloys because their response is faster than the temperature-induced responses of the conventional shape memory alloys. The difference in the sum of the magnetocrystalline anisotropy energy and the Zeeman energy for the different variants in the martensite phase in principle decides the growth of some twin variants at the expense of the others leading to net shape change. [9] If the magnetization is saturated, the magnetocrystalline energy of the twin variants alone matters as the Zeeman energy is the same for all the variants in this case. However, low twinning stress is a prerequisite for the rearrangement of the variants. Thus the condition for MFIS to occur can be written as: $K > \varepsilon_0(\sigma_{tw} + \sigma_{ext})$, where K is the magnetocrystalline anisotropy, $\sigma_{tw}$ is the twinning stress (minimum stress required to move twin boundaries), which should be ideally low for MFIS to occur through twin boundary motion, $\sigma_{ext}$ is the external stress which must be less than the critical or blocking stress and $\varepsilon_0$ is the maximum strain.[7]

Amongst the FSMAs, the stoichiometric composition $Ni_2MnGa$ in the Ni-Mn-Ga alloy system is an important intermetallic compound since it exhibits the largest MFIS (~10%) and shows other multifunctional properties like large negative magnetoresistance and magnetocaloric effect as well.[ 10, 11] The stoichiometric $Ni_2MnGa$ shows martensite (structural) and magnetic transitions at 210K and 370 K, respectively.[1,12] The large magnetic field induced strain in Ni$_2$MnGa is associated with the presence of modulated structure in the low temperature martensite phase,[1,2,3] arising from shuffling of the (110) planes along the [1-10] directions.[13, 14] Hence the presence of modulation in the crystal structure of the martensite phase is an important indicator of the high magnetic field induced strain in Ni-Mn-Ga ferromagnetic shape memory alloys.[3, 15]

Ni$_2$MnGa has a cubic L2$_1$ structure with Fm$\bar{3}$m space group in the austenite phase at room temperature. [12] On cooling below T$_{PM}$ =260K, Ni$_2$MnGa exhibits a premartensite phase followed by the martensite transition at T$_M$ = 210K. The premartensite phase has also got a modulated structure. [13, 16, 17] Its formation is linked with the softening of transverse acoustic TA2 phonon mode at $q$~0.33[1 1 0]. [16, 18, 19]. As a result of anomalously small TA2 soft-mode frequency at the premartensite transition temperature, diffuse x-ray scattering and minima in the thermal expansion coefficient and elastic modulus are observed.[20, 21] The premartensite transition is of first order type and originates from the magnetoelastic coupling between the magnetization and the anomalous TA2 phonon. [22] The phonon softening at premartensite transition temperature (T$_{PM}$) in Ni$_2$MnGa is very significant as compared to that reported for Ni-Al and Ni-Ti-Fe shape memory alloys.[23, 24] In neutron powder diffraction studies of the premartensite phase of Ni$_2$MnGa, its modulated structure was reported to be commensurate in the orthorhombic Pnnm space group with a three layer periodicity[13, 16-18], whereas a recent low energy electron diffraction (LEED) study suggests that the modulation may be incommensurate. [25]

The modulated structure of the martensite phase of $Ni_2MnGa$ has been extensively studied by different measurement techniques. [12, 13, 14, 17, 26, 27, 28, 29] In an early work, Martynov *et al.* reported a five-layer modulated martensite structure on the basis of single crystal x-ray diffraction (XRD) study. [26, 27] In subsequent neutron and x-ray powder diffraction studies on the structure of the martensite phase by Brown et al and Ranjan et al [13, 28], respectively, using Rietveld refinement technique, a commensurate modulation in the orthorhombic Pnnm space group with seven layer periodicity was reported. In contrast, using superspace group analysis of low resolution laboratory source x-ray powder diffraction data, Righi et al. reported an incommensurate modulation with a modulation vector q= 0.4248c* and claimed its rational approximant as a five layer modulated structure [14]. However, single crystal electron and neutron diffraction patterns reveal six satellites between the fundamental austenite reflections in the martensite phase of $Ni_2MnGa$ suggesting an approximately seven-layer modulation. [29] An incommensurate modulation with approximately seven layer modulation has also been reported in a neutron single crystal diffraction study of $Ni_2MnGa$. [17] It is important to mention here that while 3M, 5M, 7M type notations are commonly reported in the literature, the correct notations are 6M, 10M and 14M, as for Heusler structure the periodicity is completed after even number of layers. For B2 type ordering of the austenite phase, the periodicity is completed after odd number of layers only. We shall, however, continue to use the 3M, 5M, 7M notations for $Ni_2MnGa$, as most authors in the literature still use this notation only on the basis of the number of the satellite spots between the neighbouring austenite peaks of the $L2_1$ type ordered austenite phase.

We present here a brief overview of our recent work [30, 31] on the Le-Bail and Rietveld refinements of the premartensite and martensite phases of $Ni_2MnGa$ using high resolution synchrotron x-ray powder diffraction (SXRPD) patterns. Our results clearly rule out the commensurate structural modulation models [13, 28] for both the premartensite and martensite phases of $Ni_2MnGa$ and confirm their incommensurate nature. The implications of these findings in resolving the controversies about the rational approximant structure and the mechanisms of modulation are also briefly outlined. [13, 14, 32, 33, 34]

## 2. Signatures of premartensite and martensite transitions:

The signatures of the premartensite and martensite phase transitions are clearly observed in various physical properties measurements as well as diffraction patterns. For example, we show in Fig. 1 from ref [28] the temperature dependence of ac-magnetic susceptibility ($\chi$) for stoichiometric $Ni_2MnGa$ composition. Viewed from the high temperature side, the sharp increase in $\chi$ at $T_c$=360 K in this figure corresponds to the ferromagnetic transition. The signature of premartensite transition is observed as a small dip in $\chi$ near 250 K. At martensite start transition ($M_s$= 203K), $\chi$ decreases sharply due to a large increase in the magnetocrystalline anisotropy. Both the premartensite and martensite transitions show thermal hysteresis between heating and cooling cycles confirming first order nature of these transitions.

The modulation in the low temperature phases arises due to shuffling of (110) austenite planes in the [1 -1 0] direction of the Bain-distorted body centered tetragonal structure. [13] Different periodicities of the shuffling lead to **n** layer modulation usually represented as nM [13] A schematic representation of relation between cubic austenite and **n**M modulated structures of premartensite (**n**= 3) and martensite (n= 5 or 7) is shown in Fig. 2. The setting of crystallographic axes has been chosen in such a way that **c**$_M$, **a**$_M$ and **b**$_M$ are parallel to [1 -1 0], [1 1 0], and [0 0 1] directions, respectively of the cubic austenite unit cell. [14, 35]  As mentioned in the introduction section, the true periodicity of the so-called 3M, 5M and 7M structures in Heusler alloys is 6, 10 and 14 as can be seen from Fig. 2 also. For a B2 ordered structure, the periodicity remains 3, 5 and 7.

The modulated structure of the premartensite and martensite phases can be directly captured in single crystal x-ray, neutron and electron diffraction patterns through the appearance of extra reflections (satellites) below T$_{PM}$ and T$_M$ in certain reciprocal space sections (Fig.3). The number of these satellite reflections provides information about the periodicity of the modulation. The **n**M modulated structure is likely to show (**n**-1) satellites between the main reflections as schematically shown in Fig.3 for the 3M and 7M commensurate modulations. Thus commensurate 3M modulation leads to two satellites (Fig. 3b) between the main (2 0 0) and (2 0 3) reflections whereas 7M modulation shows six satellites between (2 0 0) and (2 0 7) main reflections (Fig.3c). It is because of this, the 3M, 5M and 7M notation continues to be used for the Heusler structure also even though the real space periodicity gets doubled. All the satellite peaks may not be always observable due to space group extinctions. If the modulation is incommensurate, the satellite peak positions may be unevenly spaced.

The structural signatures of the two transitions are clearly observed in powder diffraction patterns also. Fig. 4 compares the high resolution synchrotron x-ray powder diffraction profiles of the austenite, premartensite and martensite phases of Ni$_2$MnGa at 300 K, 230 K and 90 K respectively, around the cubic (220) austenite peak position. The appearance of a satellite peak in the 230 K pattern is the structural signature of the modulated structure of the premartensite phase. New reflections resulting from the splitting of the cubic (220) peak and appearance of additional satellite peaks in the 90 K pattern is the characteristic feature of the modulated structure of the martensite phase.

## 3. Structure of the premartensite phase

a) **Commensurate model:**

Using medium resolution neutron powder diffraction patterns, Brown et al. [13] performed Reitveld refinement of the premartensite phase considering an orthorhombic structure (space group Pnnm) with 3*M* commensurate modulation. However, Le-Bail refinements carried out by us [30] using high resolution synchrotron x-ray powder diffraction patterns for the same space group and 3M modulation led to pronounced mismatch between the observed and calculated peak positions for the commensurate model, eventhough the refined lattice parameters (a= 4.11499(2) Å, b= 12.34530(8) Å, and c= 5.81955(1) Å) obtained from Le-Bail refinement are in good agreement with those given by Brown et al. [13]. Figs. 5 (b) and (c) depict the Le-Bail fits between the calculated and observed profiles for some of the main and satellite peaks of the premartensite phase, respectively, on an expanded scale [30]. It is evident from these fits that the commensurate model accounts fairly well for the main peaks. However, there is a distinct misfit between the observed and calculated peak positions for the satellite reflections (Fig. 5(c)). This misfit cannot be modeled by using a different periodicity of the commensurate modulation and indicates the failure of the commensurate modulation model.

b) **Incommensurate model :**

The failure of the commensurate modulation model necessitates to revisit the modulated structure of the premartensite phase using super space group approach for incommensurate modulations. Although incommensurate modulation of the premartensite phase has been earlier reported by Fukuda *et al.* [33] and Kushida et al [34] on the basis of neutron diffraction measurements performed on single crystals of Ni$_2$MnGa, the structure refinements were carried out assuming commensurate modulation only. Therefore, they could not determine the correct atomic positions in the incommensurate premartensite phase. [33, 34] For incommensurate modulation of Ni$_2$MnGa, the real crystal structure and atomic positions can be obtained by carrying out structure refinements in the (3+1) D superspace using software packages like Jana2006. [36] In this approach, the powder diffraction pattern is divided into two parts: (i) main reflections corresponding to the basic structure having higher intensities and (ii) satellite reflections corresponding to the modulation having comparatively weaker intensities. All the main reflections could be indexed using an orthorhombic cell with space group *Immm* and unit cell parameters $a$=4.114889(14) Å, $b$=5.819336(27) Å, and $c$=4.115351(18) Å following the convention of Ref.14 for choosing the crystallographic axes. The superspace group *Immm* (00γ) s00 successfully accounts for the entire powder diffraction profile including both the main and the satellite peaks [30]. If one assumes commensurate modulation with a modulation wave vector of *q*= (0, 0, 1/3), as in the model of Brown et al. [13] and Kushida et al. [34], the calculated satellite peak positions are found to be considerably shifted from the observed positions, similar to the shifts shown in Fig 5(c) for the refinements employing the *3M* modulation and Pnnm space group. On allowing the refinement of the modulation wave vector *q*, the Rietveld fits are found to become excellent, as can be seen from Fig. 6(a). Both the main and the satellite reflections are very well accounted using the incommensurate modulation model for a modulation wave vector $q= 0.33761(5)c^*$, as can be seen from Figs. 6(b) and 6(c) which depict the Rietveld fits for a few selected main and satellite reflections on an expanded scale. The structural parameters obtained by super space Rietveld refinement for the incommensurate model are given in Table 1. The fact that $q$ is slightly different from the commensurate value of $(1/3)$ $c^*$ confirms the incommensurate nature of modulation in the premartensite phase of $Ni_2MnGa$ with a modulation vector $q= (1/3 +\delta)$ $c^*$, where $\delta= 0.00428$ is the degree of incommensuration of the structure.

## 4. Structure of the Martensite phase

**a) Commensurate model.**

As mentioned in the introduction section, the first Rietveld refinement of the modulated structure of the martensite phase was carried out using medium resolution neutron powder diffraction data by Brown et al. [13] taking an orthorhombic Pnnm space group with 7M commensurate modulation. Ranjan et al. also supported commensurate 7M modulation and Pnnm space group based on Rietveld refinements using medium resolution x-ray powder diffraction data from a rotating anode based x-ray source. [28]

To check the validity of the commensurate structural modulation in the martensite phase of $Ni_2MnGa$ using high resolution synchrotron x-ray powder diffraction patterns recorded at 90 K, LeBail refinement has been carried out recently [31]. The results of refinements, using the commensurate 7M modulation and *Pnnm* space group proposed by Brown et al. and Ranjan et al. [13, 28]**,** shown in Fig. 7 reveal clear mismatch between the observed and calculated peak profiles for several Bragg reflections which suggests that the nature of modulation in the martensite phase of stoichiometric $Ni_2MnGa$ is not commensurate. It was possible to arrive at such a conclusion on account of the use of high resolution x-ray powder diffraction

**b) Incommensurate model.**

The incommensurte modulated structure of the low temperature martensite phase of $Ni_2MnGa$ has been reported using different diffraction techniques (x-ray, neutron and electron diffraction) [13, 29, 33, 34, 37, and 38]. Righi *et al.* have performed Rietveld refinement in the (3+1) dimensional superspace using laboratory source x-ray powder diffraction data.[14] They obtained an incommensurate modulation with a modulation vector $q$ = 0.4248(3) $c^*$ = (2/5 +0.0248) $c^*$ and concluded that the nearest rational approximant to the incommensurate phase is of 5*M* type. However, subsequent single crystal diffraction studies [29] have revealed that the incommensurate phase exhibits six spots between the fundamental reflections suggesting a 7*M* type rational approximant for the stoichiometric $Ni_2MnGa$ composition, which is in disagreement with Righi et al.'s conclusions. This controversy has been resolved recently by Rietveld refinement in (3+1)-D superspace using high resolution synchrotron x-ray powder diffraction data. [31]

As in the case of the premartensite phase, the diffraction pattern of the martensite phase recorded at 90K could be divided into two groups: the main reflections and the satellite reflections. Following Righi et al., all the main reflections corresponding to the basic structure of $Ni_2MnGa$ could be indexed well using the orthorhombic space group *Immm* with unit cell parameters a= 4.21853(2) Å, b=5.54667(2) Å and c=4.18754(1) Å. LeBail refinement using the super space group *Immm(00γ)s00* assuming the commensurate value of the modulation vector [q= 3/7 $c^*$] and first [31] harmonics of the sinusoidal modulation was unable to fit the satellite reflections. The calculated peak positions were found to be shifted away from the observed ones. [31] Use of incommensurate value of the modulation vector reported in Ref. [12] (*q= 0.4248)* led to even worse fit compared to that for the commensurate modulation with [q= 3/7 c*].[31] This indicates that the modulation vector, as reported by Righi et al. [14], cannot account for the satellite peak positions precisely. Refinement of the modulation vector led to significantly better fit between the observed and the calculated peak positions with an incommensurate value of the modulation vector $q$= 0.43160(3) $c^*$. [31] However, it was found that consideration of only the first order harmonics (i.e., first order satellite reflections (h k l ± 1)) in the refinements as per Righi et al. [14] did not account for a number of weak reflections with rather low intensities. These reflections were confirmed to be second and third order satellites (hk l ±2) and (hk l+3) by carrying out LeBail and Rietveld refinements using third order harmonics. It is important to point out here that the observation of second and third order satellites in the powder diffraction pattern of the martensite phase of $Ni_2MnGa$ could become possible for the synchrotron data only. Using such a high resolution data, it was also found [31] that the anisotropic peak broadening functions in terms of Stephens [39] formalism, widely used for 3D periodic crystals, cannot fully account for the peak broadening of satellite reflections because of the additional broadening of the satellite peaks due to phasons [40]. The fit between the observed and calculated peak profiles improved significantly on use of a fourth-rank covariant strain tensor based formalism with a distribution of the strain-tensor components implemented in JANA2006 [36]. This was the first experimental evidence of phason broadening in the powder diffraction pattern of $Ni_2MnGa$ resulting from the fluctuations in the incommensurate modulation vector [31] although phasons associated with CDW have been reported earlier in inelastic neutron scattering studies.[41]

The observed and calculated profiles obtained by Rietveld refinements taking into account third order satellites and phason broadening are shown in Fig.8. [31] Excellent fit for the second and third order satellites using the incommensurate modulation model is illustrated in the inset of this figure. The refined basic cell parameters a= 4.21861(2) Å, b= 5.54696(3) Å, and c= 4.18763(2) Å and modulation vector $\mathbf{q}$= 0.43160 (3)$c^*$ are close to the values obtained by LeBail refinement. The refined atomic positions and the amplitudes of the modulation functions of the incommensurately modulated phase are listed in Table 2.

## 5.  Concluding Remarks

Structure refinements of the modulated structure of the premartensite and martensite phases of stoichiometric Ni$_2$MnGa ferromagnetic shape memory alloy in the (3+1) D superspace using high resolution synchrotron x-ray powder diffraction data confirms incommensurate nature of modulation. Evidence for the phason broadening of the satellite peaks due to fluctuations in the modulation wave vector further confirms the incommensurate nature of the modulation. The modulation wave vectors are found to be $\mathbf{q}$= 0.33761(5)$c^*$= (1/3+δ)$c^*$ [30] and 0.43160(3)$c^*$= (3/7+δ) $c^*$ [31] for the premartensite and martensite phases, respectively.

The amplitude of modulation for different atoms can be used to distinguish between the adaptive phase and displacive modulation models proposed in the literature for the martensite phase of Ni$_2$MnGa. [35] In the adaptive phase model [32], the modulated structure is visualized as a nanotwinned state of the Bain distorted phase, which maintains the invariance of the habit plane between the austenite and the martensite phases. In the soft phonon model, the origin of modulation has been related to a TA2 soft phonon mode in the transverse acoustic branch along [110] direction of the austenite phase [16, 17, 20], which has been supported by the observation of change in the modulation period leading to a premartensite phase before the final structural transition to the martensite phase [20]. The amplitudes of displacive modulations for the different atomic sites are required to be identical for the adaptive phase model but may be dissimilar also for the soft mode models. [35] Since the amplitudes of modulation for the different atomic sites and the directions of the atomic displacements are considerably dissimilar for $Ni_2MnGa$ (see Table 2), the adaptive phase model may be rejected. This leaves the soft phonon model as the most plausible model for modulation in the martensite phase of $Ni_2MnGa$. [35] The observation of charge density wave in $Ni_2MnGa$ also indicates that the modulation is driven by soft- phonon modes. [25]

Finally, we comment on the controversy about the 5M and 7M like modulations in the martensite phase of $Ni_2MnGa$. For this we show in Fig.9 the simulated single crystal diffraction patterns obtained using refined parameters given in Table 2 and including the satellites upto the 3$^{rd}$ order in the simulations [31]. There are clearly six satellites between the main reflections in this figure and they are unevenly spaced due to incommensurate nature of modulation. This is in good agreement with the electron diffraction patterns of Fukuda et al. [29]. The presence of six satellite spots in Fig.9 clearly confirms 7M like modulation and rejects the 5M like modulation reported earlier by Righi et al. on the basis of simulations considering upto the second order satellites only.

## Acknowledgements:

We thank our collaborators V. Petricek, P. K. Mukhopadhyay, A. Hill and P. Rajput. S. S. thanks Alexander von Humboldt foundation, Germany and CSIR, India for Fellowships. DP thanks the Science and Engineering Research Board (SERB) of India for the award of the J.C. Bose National Fellowship.

**Figure captions:**

Fig.1: Temperature variation of ac susceptibility of $Ni_2Mn_{1.05}Ga_{0.95}$ during heating and cooling. Figure has been taken from [Ref. 28]

Fig.2: Schematic representation of unit cell orientation correlation between austenite, body centered tetragonal (bct) martensite, 3M premartensite, 5M martensite and 7M martensite superstructures. M and A correspond to the martensite (or premartensite) and austenite, respectively.

Fig.3: Schematic representation of single crystal diffraction pattern for (a) Austenite $L2_1$ (b) Premartensite 3M superstructure and (c) 7M martensite superstructure. Black arrows are for main reflections and green are for extra reflection due to formation of superstructure.

Fig.4: The (220) austenite peak region of $Ni_2MnGa$ at (a) 300 K (austenite phase), (b) 230 K (premartensite phase) and (c) 90 K (martensite phase). The satellite peaks are marked with asterisk. The satellite peaks correspond to premartensite and martensite phases have been marked as PM and M, respectively.

Fig.5. (a) Le Bail fitting (red line) of the high resolution SXRPD pattern (black dots) of $Ni_2MnGa$ at 230 K (premartensite phase) using the commensurate 3M model. The residue is shown as green solid line. The blue ticks represent the Bragg peak positions. Inset shows the satellite reflections marked by arrows. Raw data (open circles) and the fitted curve (red line) for (b) the main reflections and (c) the satellite reflections. Arrow in (c) indicates the satellite reflection. Figure obtained from Ref. 30.

Fig. 6. (a) Rietveld fitting (red line) of the high resolution SXRPD pattern (black dots) of the premartensite phase of $Ni_2MnGa$ at 230 K using an incommensurate (3+1) D model. The inset shows the indexing of the satellite reflections. Raw data (open circles) and the fitted curve (red line) for (b) the main reflections and (c) the satellite reflections. Arrow in (c) indicates the satellite reflection. Here the peaks have been indexed with four integers (h k l ±m), where m (= 1) is the order of satellite reflections. Figure obtained from Ref. 30.

Fig.7. Le Bail fitting of the XRD pattern in the martensite phase (90K) of $Ni_2MnGa$ using commensurate 7M model. The experimental data, fitted curves and the residue are shown by black circles, red line and green line, respectively. The blue ticks represent the Bragg peak positions. Inset show the fitting for some satellite peaks.

Fig. 8. Rietveld fitting for the incommensurate modulated martensite phase of $Ni_2MnGa$ at 90K. The experimental data, fitted curve and the residue are shown by dots (black), continuous line (red) and bottom most plot (green), respectively. The tick marks (blue) represent the Bragg peak positions. The right inset shows main region in expanded scale. The inset on the left shows the fit for $2^{nd}$ and $3^{rd}$ order satellites. Here the peaks have been indexed with four integers (h k l ±m), where m (= 1, 2 and 3) are the order of satellite reflections. [31]

Fig.9. Simulation of the (010) section of the reciprocal space of the 7M incommensurate modulated structure using satellite peaks up to $3^{rd}$ order. Red spots are the main reflections and Green spots are satellites. Numbers 1, 2 and 3 represent the order of the satellites. [31]

Fig.1

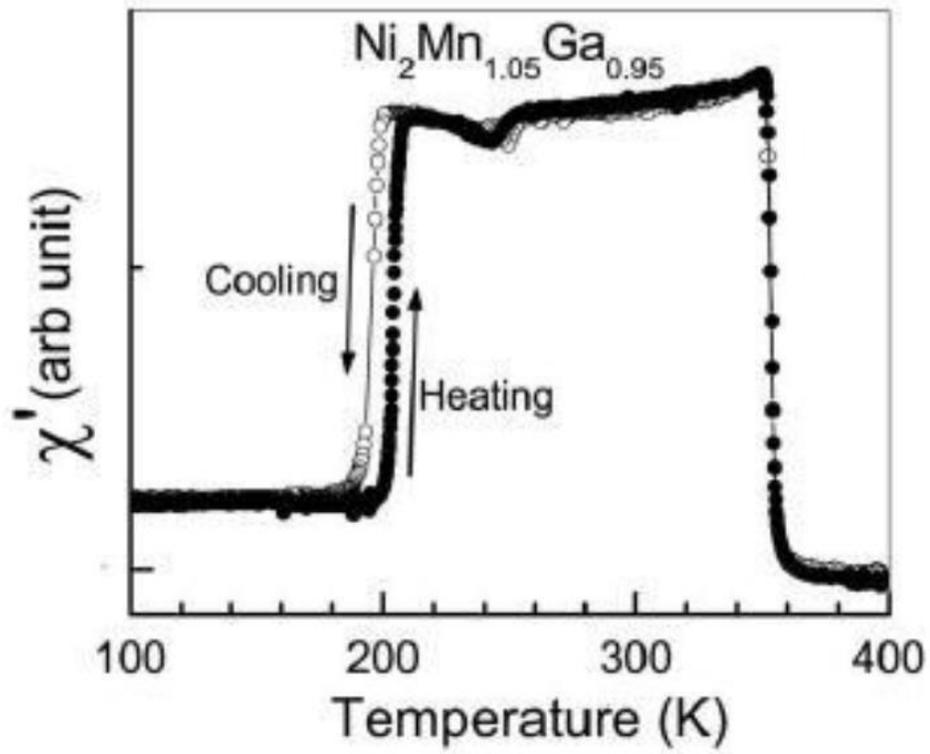

Fig.2

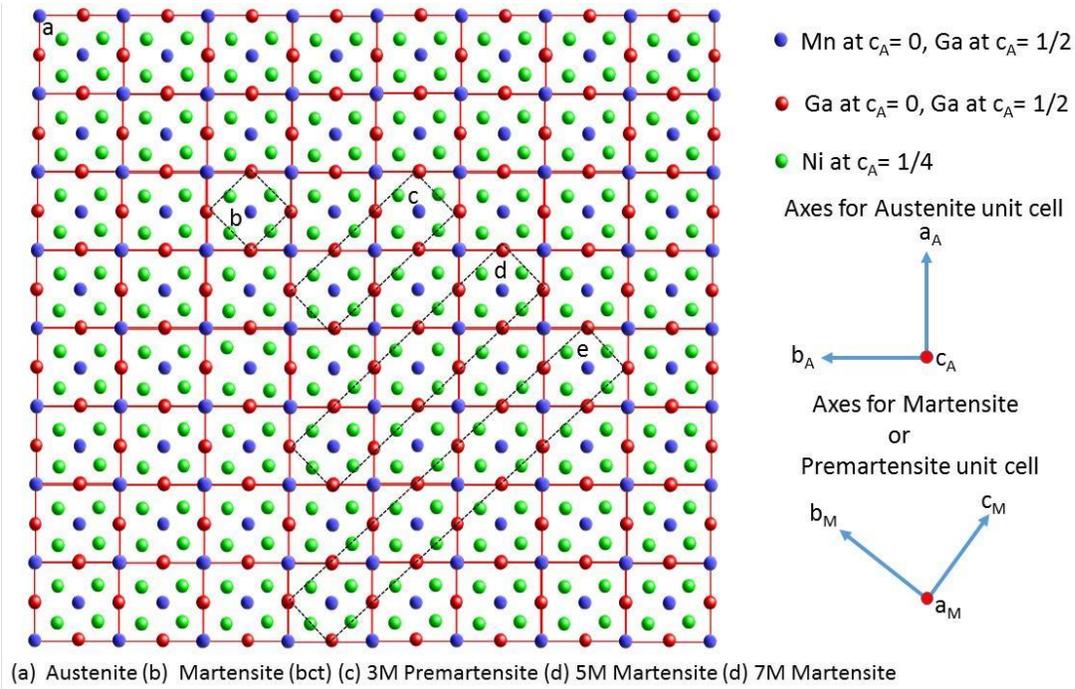

(a) Austenite (b) Martensite (bct) (c) 3M Premartensite (d) 5M Martensite (d) 7M Martensite

Fig.3

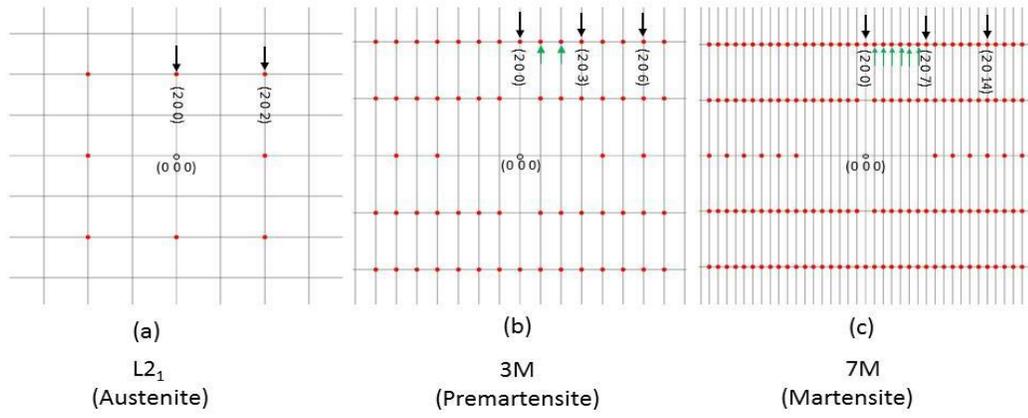

(a) L2$_1$ (Austenite)

(b) 3M (Premartensite)

(c) 7M (Martensite)

Fig.4

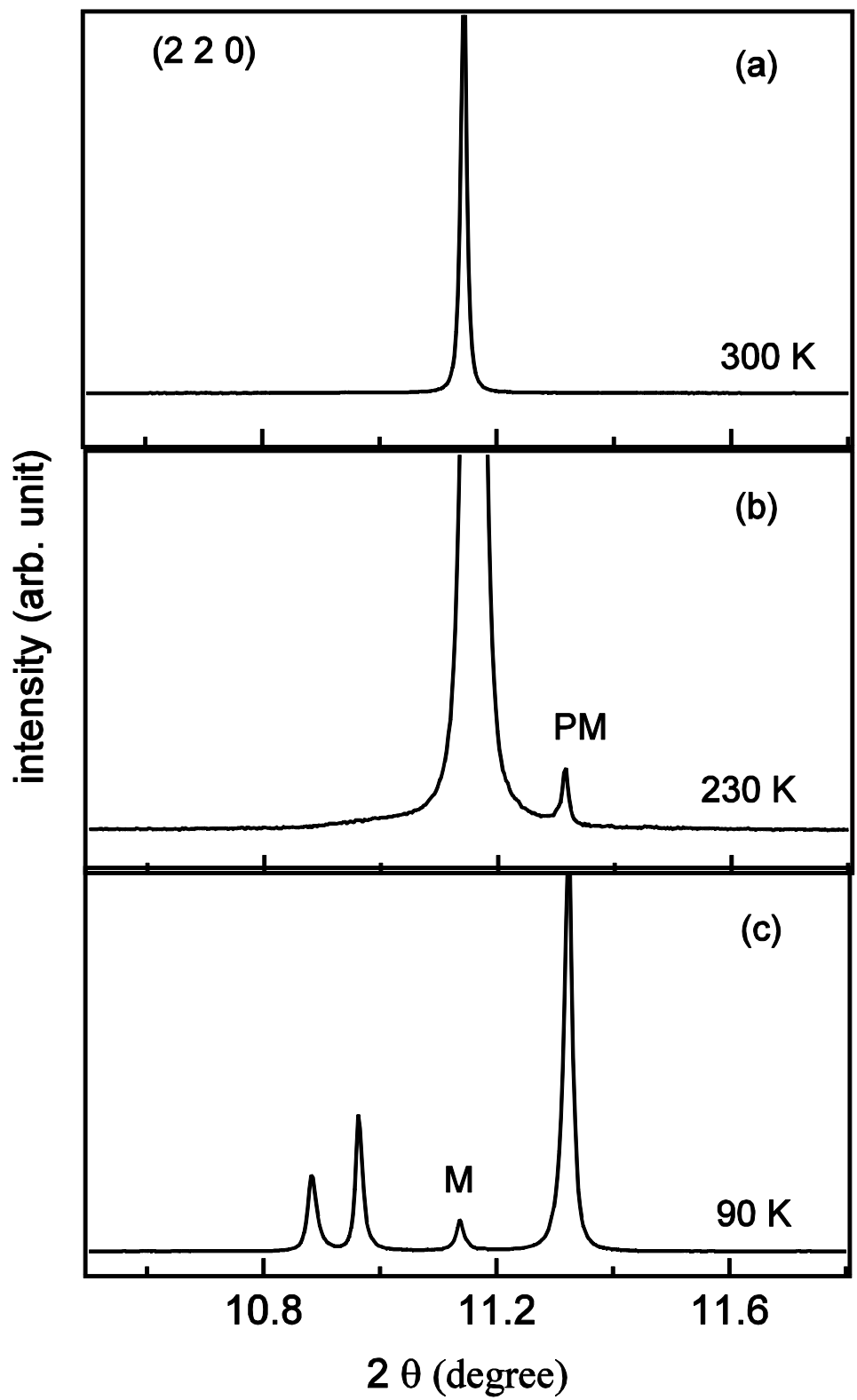

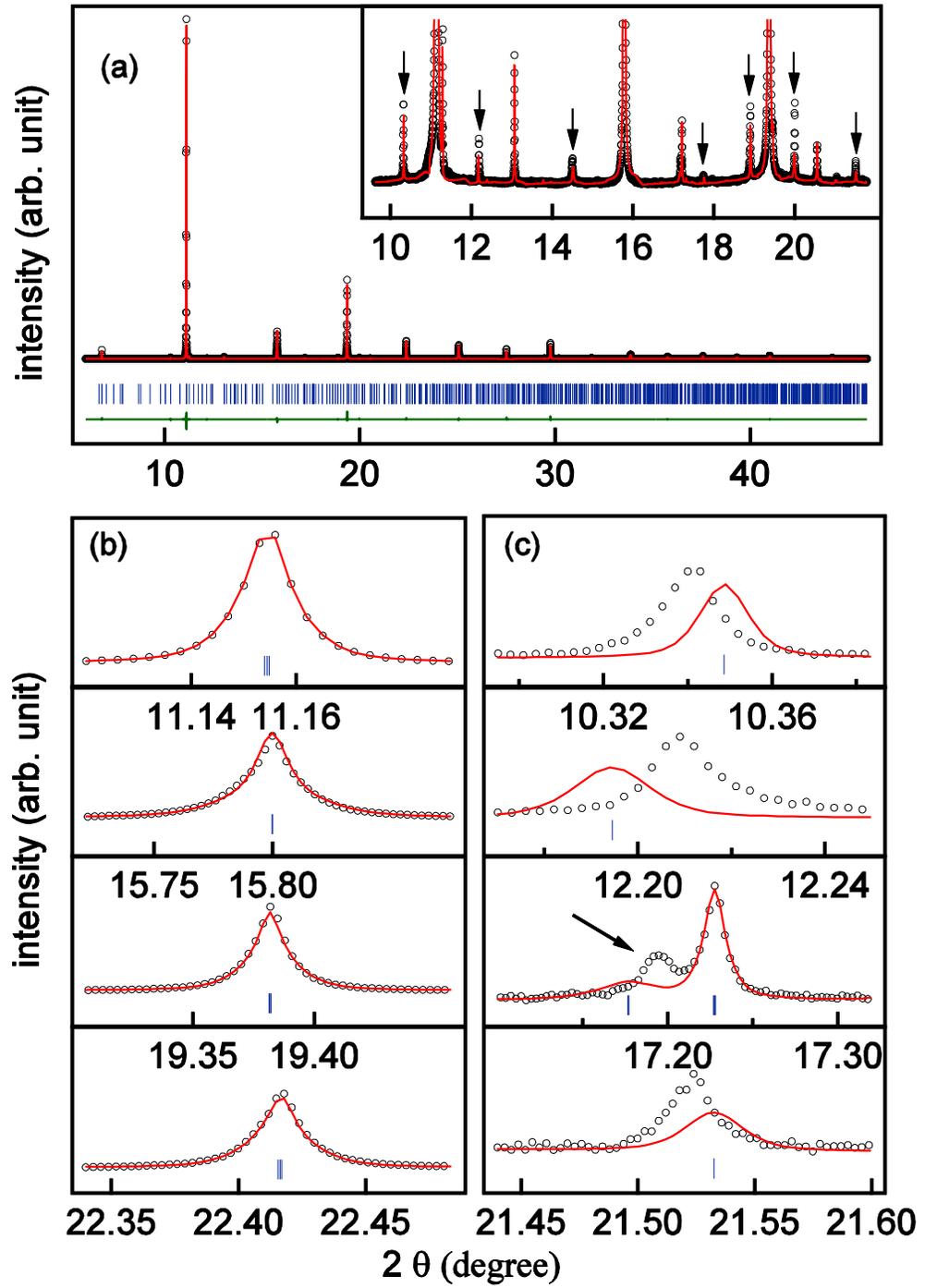

Fig.6

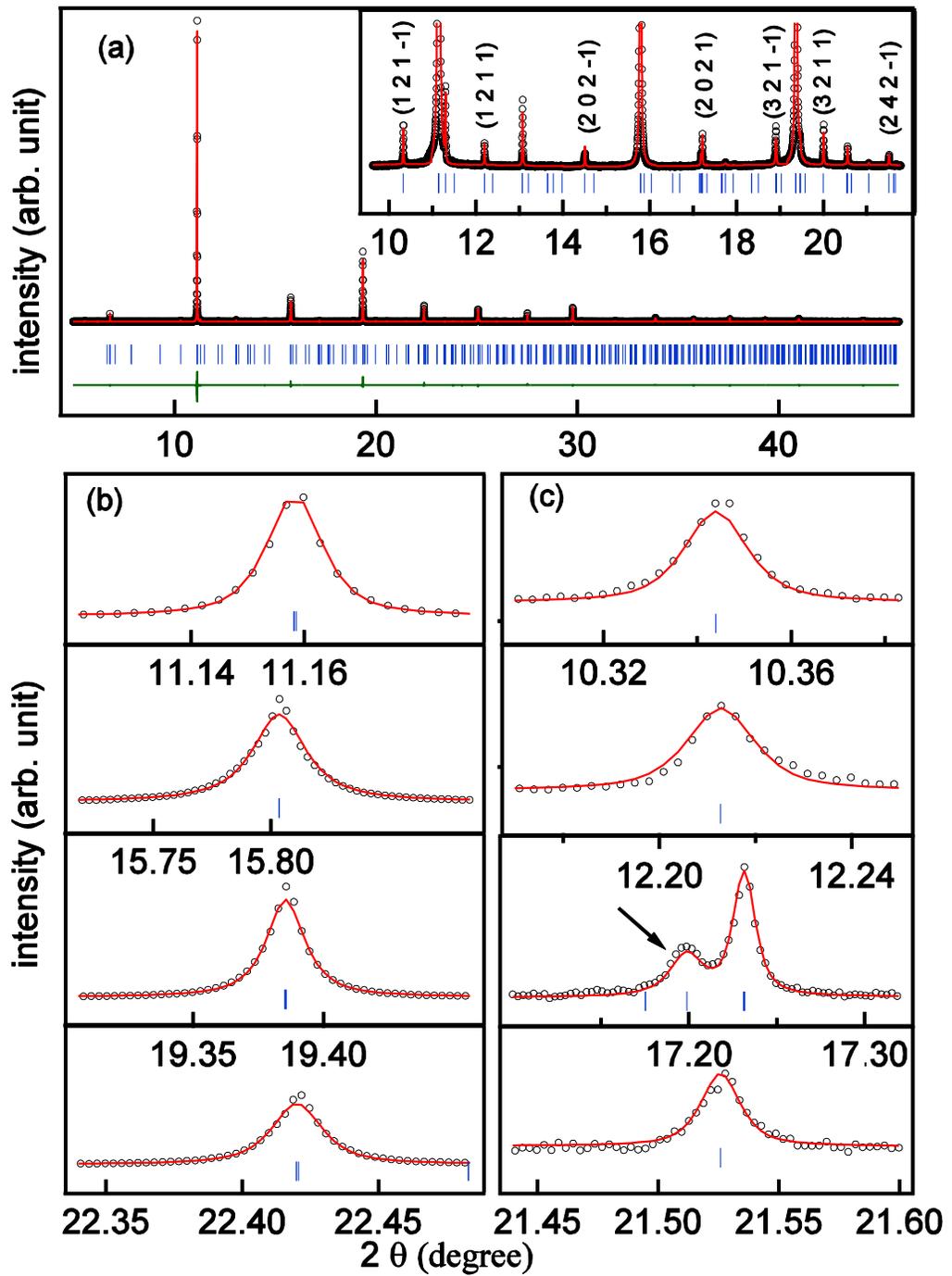

Fig.7

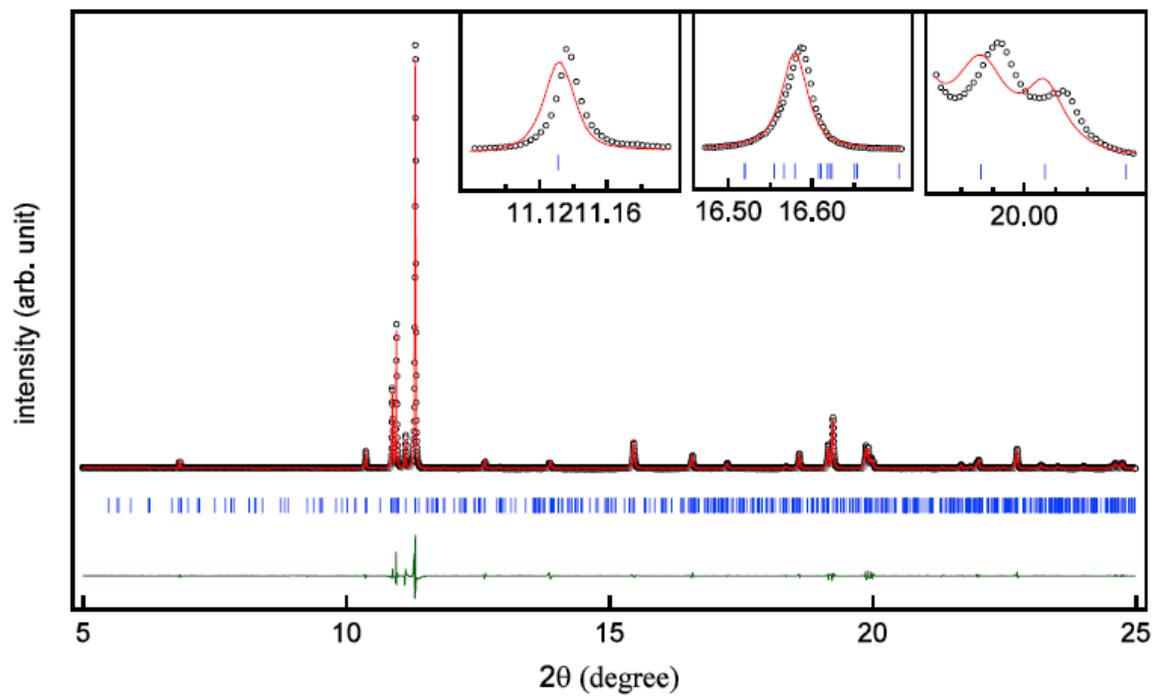

Fig.8

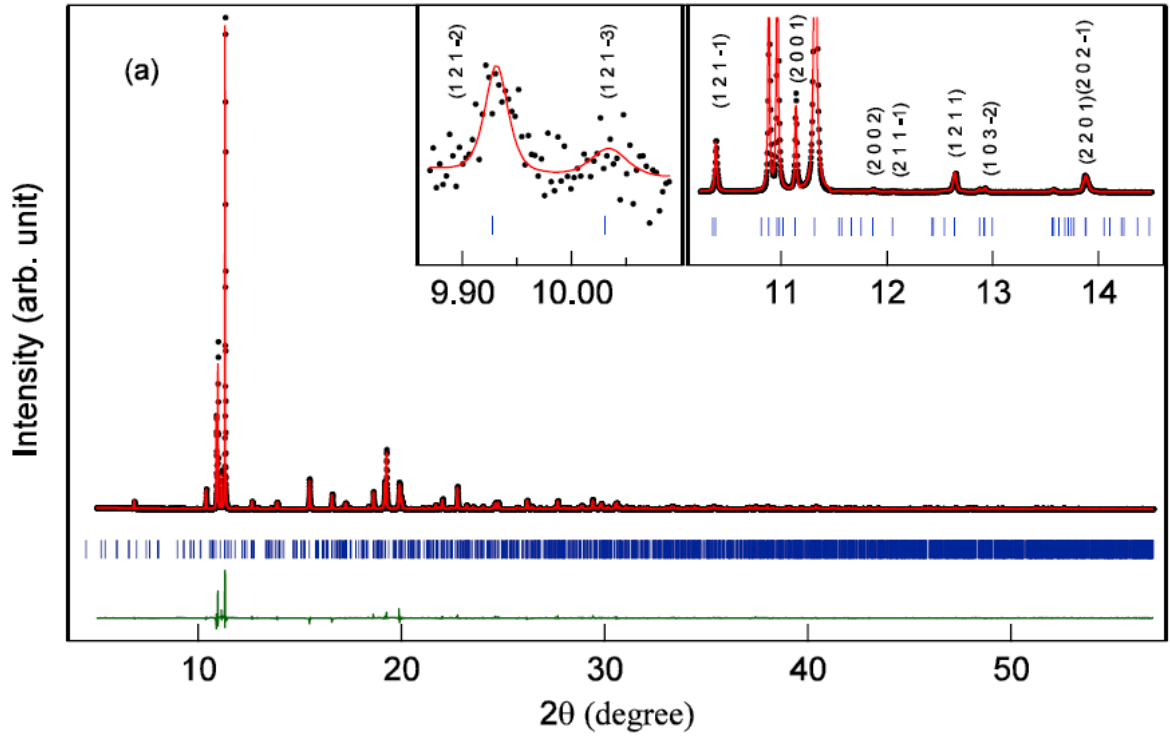

Fig.9

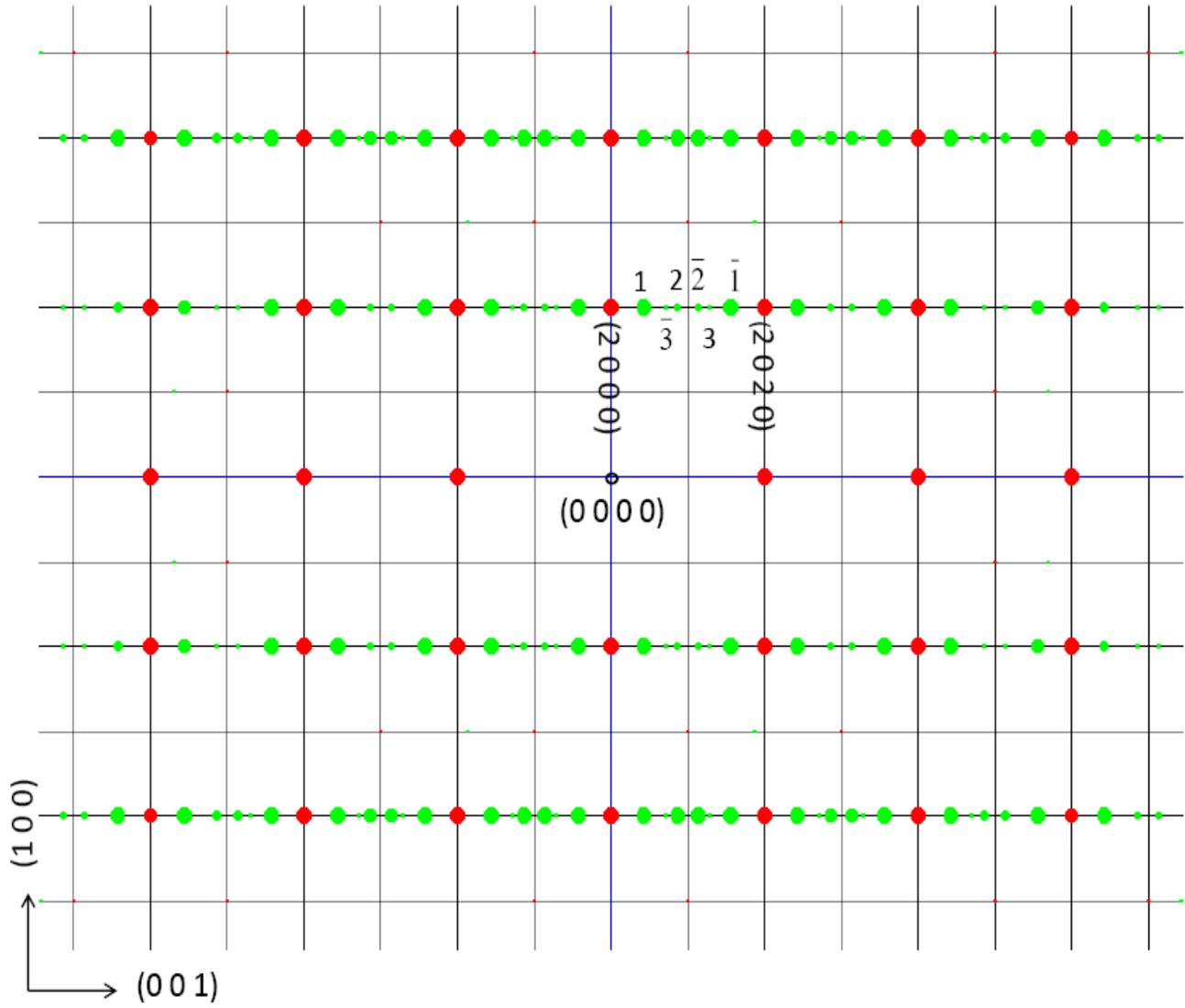

Table 1: Atomic positions ($x, y, z$), atomic displacement parameter ($U_{iso}$) and amplitudes ($A_1$, $B_1$) of the modulation function of the incommensurate modulated premartensite phase of Ni$_2$MnGa. The values of $A_1$ and $B_1$ are in Å. Table reproduced from Ref.30.

| Atom | Wyckoff position | Modulation amplitude | x | y | z | $U_{iso}$ (Å$^2$) |
|---|---|---|---|---|---|---|
| Ga1 | 2d | | 0 | 0.5 | 0 | 0.0043(2) |
| | | $A_1$ | 0.0227(7) | 0 | 0 | |
| | | $B_1$ | 0 | 0 | 0 | |
| Mn1 | 2a | | 0 | 0 | 0 | 0.0036(3) |
| | | $A_1$ | 0.0184(10) | 0 | 0 | |
| | | $B_1$ | 0 | 0 | 0 | |
| Ni1 | 4h | | 0.5 | 0.25 | 0 | 0.0047(3) |
| | | $A_1$ | 0.0246(6) | 0 | 0 | |
| | | $B_1$ | 0 | 0 | 0 | |

**Table 2:** Atomic positions $(x, y, z)$, atomic displacement parameter $(U_{iso})$ and amplitudes $(A_1, B_1, A_2, B_2, A_3, B_3)$ of the modulation function of the incommensurate modulated martensite phase of $Ni_2MnGa$. Table reproduced from Ref. 31.

| Atom | Wyckoff position | Modulation amplitude | x | y | z | $U_{iso}$ (Å$^2$) |
|---|---|---|---|---|---|---|
| Ga1 | 2d |  | 0 | 0.5 | 0 | 0.0037(5) |
|  |  | $A_1$ | 0.0657(9) | 0 | 0 |  |
|  |  | $B_1$ | 0 | 0 | 0 |  |
|  |  | $A_2$ | 0 | 0 | 0.001(3) |  |
|  |  | $B_2$ | 0 | 0 | 0 |  |
|  |  | $A_3$ | -0.005(2) | 0 | 0 |  |
|  |  | $B_3$ | 0 | 0 | 0 |  |
| Mn1 | 2a |  | 0 | 0 | 0 | 0.0030(6) |
|  |  | $A_1$ | 0.0665(12) | 0 | 0 |  |
|  |  | $B_1$ | 0 | 0 | 0 |  |
|  |  | $A_2$ | 0 | 0 | -0.001(4) |  |
|  |  | $B_2$ | 0 | 0 | 0 |  |
|  |  | $A_3$ | -0.001(3) | 0 | 0 |  |
|  |  | $B_3$ | 0 | 0 | 0 |  |
| Ni1 | 4h | | 0.5 | 0.25 | 0 | 0.0013(4) |
| | | $A_1$ | 0.0618(9) | 0 | 0 | |
| | | $B_1$ | 0 | 0 | 0 | |
| | | $A_2$ | 0 | 0 | 0.000(3) | |
| | | $B_2$ | 0 | -0.0029(7) | 0 | |
| | | $A_3$ | -0.004(2) | 0 | 0 | |
| | | $B_3$ | 0 | 0 | 0 | |